\documentclass[aps,prb,amssymb,a4paper,amsmath,10pt]{revtex4} 
\usepackage{amsfonts}
\usepackage{amsmath}
\usepackage{amssymb}
\usepackage{graphicx}
\usepackage{color}
\usepackage{subfigure}
\begin{document}

\title{Spin splitting and switching effect in a four-terminal two-dimensional electron gas nanostructure}
\author{Zijiang Wang$^{1}$, Jianhong He$^{1,2}$, Huazhong Guo$^{1}$\footnote{
 Electronic mail: guohuazhong@scu.edu.cn}
} \affiliation{$^1$ Laboratory of Mesoscopic and Low Dimensional
Physics, College of Physical Science and
Technology, Sichuan University, Chengdu 610064, P. R. China\\
 $^2$
National Institute of Measurement and Testing Technology, Chengdu
610021, P. R. China}
\date{\today}

\begin{abstract}
We have studied the spin-splitting effect in a four-terminal two-dimensional (2D) electron gas system with two potential barriers generated by two surface metal gates and an external perpendicular magnetic field. The calculations show that by tuning the voltage applied on the gates, the injected spin-unpolarized current can be split into different spin currents with a high efficiency. The split currents flow out of the geometry from different output leads separately. The spin freedom of the outputs can be controlled by simply tuning voltage on gates. This phenomenon is a result of the combination of three effects - the potential barriers, Zeeman splitting and edge current. Furthermore, by tuning the voltage on gates, the outflow spin of current in one terminal can be switched. Therefore, these features open up a possibility for making a spin filter or a switcher device by applying the four-terminal 2D electron gas system.
\end{abstract}

\maketitle

The manipulation of spin is a significant issue for spintronics. In the spin transistor proposed by Datta and Das \cite{spintransistor}, spin-polarized electrons are injected into the semiconductor from a ferromagnet, and by tuning the spin-orbit interaction (SOI), the spin of electrons is manipulated in a desired manner. However, the conductivity mismatch between the semiconductor and ferromagnet leads to a low efficiency of the spin injection, less than 0.1\% \cite{spininjector}. To solve this problem, one approach is to inject the spin-polarized current from the magnetic semiconductor \cite{magneticleads}. By using this method, the injection efficiency of the device is reported to be 90\% \cite{efficientinjector}. Another approach is to use the SOI to achieve spin filters without ferromagnets. Several spin filters were proposed by using this method, e.g., quantum dots \cite{QDspinfilter}, semiconductor nanowires \cite{QWspinfilter}, quantum point contacts (QPC) \cite{QPCspinfilter}.

Recently an antidot in a quantum wire was proposed as a spin filter/switcher \cite{antidot}. The operating method to control the spin freedom is based on the effect of resonant backscattering from one edge state into another through localized quasibound states, combined with the effect of Zeeman splitting of the quasibound states in adequately high magnetic field. The author of Ref. 8 expected that by varying parameters of the system, e.g., the magnetic field, the Fermi energy, the antidote size, the conductance of the device could be tuned such that the current in one lead could be due exclusively to spin-up or spin-down electrons.

In this letter, we propose a method to create a spin splitter/switcher. To split different spin current, we need at least three terminals, one for the input current, and another two terminals for different output spin currents. From experimental point of view, varying gate voltage is more convenient than varying magnetic strength. In order to control the spin currents of the three outputs, we utilize two gates on two the conduction channels. The existence of edge current arising from Landau levels has been testified in a semiconductor heterostructure \cite{edgecurrent}. In the ballistic transport regime, we use the gate-controlled potential barriers, combined with Zeeman splitting effect and edge current to control the spin freedom of the outputs. In a word, we design a four-terminal nanostructure with two surface metal gates, and show that by tuning the voltage of gates, the injected spin-unpolarized current can be split into different spin current with a high efficiency, and flow out of the geometry from different terminals. Furthermore, by tuning the voltage of gates, the outflow spin can be switched.

\begin{figure}[h]
\includegraphics*[height=5cm,keepaspectratio=true,angle=0]{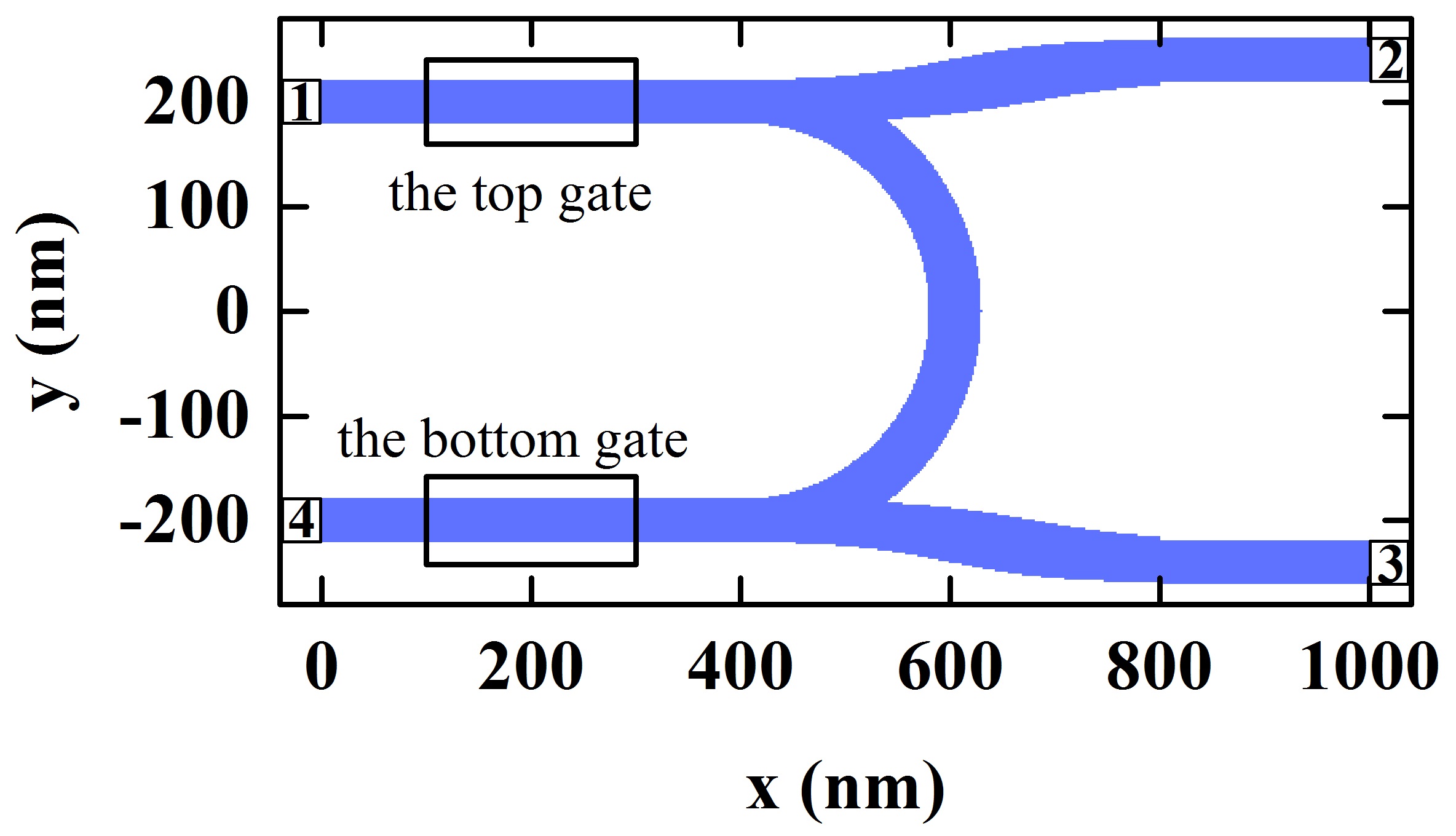}
\caption{Schematic of the four-terminal nanostructure with two rectangular gates. Four contacts are numbered from 1 to 4. The unpolarized current is injected from Contact 2.}
\label{fig1}
\end{figure}

We consider the 2D electron gas in a four-terminal system depicted in Fig. \ref{fig1}. The scattering region consists of three parts. The first part is a horizontal narrow long strip at the top of the region, which is slightly curved around $x$ $\sim$ 450nm to 800nm. The second part is a mirror-symmetrical strip about the $x$-axis. Finally the third part is a half ring for just connecting the upper and lower parts. The whole geometry is mirror-symmetrical about the $x$-axis. Contact 2 acts as the injector of spin-unpolarized electronic current and the other three contacts act as outputs.

The 2D electron gas locates at the depth of $d$ beneath the surface of the heterojunction, and two rectangle metal gates on the surface are defined by $l_1<x<r_1$, $b_1<y<t_1$ and $l_2<x<r_2$, $b_2<y<t_2$ respectively. These gates are located above the branches connected with Contact 1 and Contact 4 respectively, as depicted in Fig. \ref{fig1}. They are mirror-symmetrical about the $x$-axis. By applying voltages on these gates, we can control the electrochemical potential in the region beneath the gates.

When external magnetic field $\mathbf{B}=(0, 0, -B)$ is applied on the electron gas plane, the Hamiltonian in the system is given by:
\begin{equation}
H=\left[ \frac{\mathbf{P}^{2}}{2m^*}+U(\mathbf{r})\right]\mathbf{1}+\frac{1}{2}g\mu _B B\sigma _z
\label{eq1}
\end{equation}
where $\mathbf{P}=\hbar \mathbf{k}+e\mathbf{A}$, $\mathbf{k}=-i\nabla$, $\mathbf{A}=(-By,0,0)$ is the vector potential, $m^*$ is the effective conduction-band electron mass, $\mathbf{1}$ is the 2$\times$ 2 unit matrix, $g$ is the effective Lande factor, $\mu _B$ is the Bohr magneton, and $\sigma _z$ is the $z$th Pauli matrix. The second term in Eq. (\ref{eq1}) describes the spin-dependent Zeeman split energy. We assume that the confinement in the $z$ direction is so strong that the electrons occupy the ground state due to quantum size effect along this direction. Therefore the confinement potential energy $U(\mathbf{r})$ is given by:
\begin{equation}
U(\mathbf{r})=U_c(x,y)+U_c(z)+U_g(x,y)
\label{eq2}
\end{equation}
where $U_c(x,y)$ is the confinement potential energy in the $x$-$y$ plane, which is described by the hard-wall confinement in the $x$-$y$ plane, $U_c(z)$ is the confinement potential energy in $z$ direction, which confines electrons in the $x$-$y$ plane, and $U_g(x,y)$ is the potential energy aroused by surface gates. According to J. H. Davies’ model \cite{Daviesmodel}, in a depth of $d$, the potential $\phi (x,y,d)$ aroused by a rectangle gate defined by $l<x<r$, $b<y<t$ can be calculated analytically. For our system with two gates, we have:
\begin{equation}
U_g(x,y)=-e\phi _T(x,y,d)+\left[ -e\phi _B(x,y,d) \right]
\label{eq4}
\end{equation}
where index $T$ denotes the top gate, index $B$ denotes the bottom gate and $d$ is the depth of 2D electron gas. When the gate voltage is negative, potential barrier for electrons will be formed in the channel underneath the gate, which will block electrons from passing through the channel. In the following, We use $V_{T}$ to denote the top gate voltage and $V_{B}$ to denote the bottom gate voltage.
In our model, the leads' directions are parallel to the $x$-axis. We perform the numerical calculation within the frame of tight-binding approximation by using the Kwant package \cite{Kwant}. The Hamiltonian has to be discretized on grids $(x_\mu=\mu a, y_\nu=\nu a, \mu, \nu=1, 2, 3, ...)$ with lattice constant $a$. We introduce the discrete representation of the spinor as follows: $|\Psi(x_\mu,s_\mu,y_\nu,s_\nu)\rangle=(|\varphi^\uparrow(x_\mu,y_\nu)\rangle,|\varphi^\downarrow(x_\mu,y_\nu)\rangle)^T=|\mathbf{\Psi}_{\mu,\nu}\rangle$, where $s_\mu$ is the spin index ($\mu=\uparrow,\downarrow$), $(...)^T$ denotes the matrix transposition. The discretized Hamiltonian is then given by:
\begin{equation}
\begin{split}
H=\sum_{\mu,\nu}\left\{\left[4t+U_g(x_\mu,y_\nu)\right]\mathbf{1}+\frac{1}{2}g\mu_BB\sigma_z\right\}|\mathbf{\Psi}_{\mu,\nu}\rangle\langle\mathbf{\Psi}_{\mu,\nu}| \\
-\sum_{\mu,\nu}(t'\mathbf{1}|\mathbf{\Psi}_{\mu+1,\nu}\rangle\langle\mathbf{\Psi}_{\mu,\nu}|+t\mathbf{1}|\mathbf{\Psi}_{\mu,\nu+1}\rangle\langle\mathbf{\Psi}_{\mu,\nu}|+H.c.)
\end{split}
\label{eq5}
\end{equation}
where $t=\hbar^2/(2m^*a^2)$, $t'=t\exp(-i(e/h)By_\nu a)$.

Considering the four-terminal system, the spin dependent current of spin up ($\sigma=\uparrow$) and spin down ($\sigma=\downarrow$) electrons can be calculated by B\"{u}ttiker formula \cite{mesoscopictransport}:
\begin{equation}
I_p^\sigma=\sum_qG_{pq}^\sigma(V_p-V_q)
\label{eq6}
\end{equation}
where $p$ and $q$ are terminal index ($p$, $q=1$, 2, 3, 4), $G_{pq}^\sigma$ is the conductance of spin $\sigma$ from terminal $q$ to $p$, $V_p$ is the voltage on Contact $p$. In our model Contact 2 acts as the injector of electronic current, and other contacts act as outputs. As Contact 2 acts as the injector of electrons, $V_2<0$. We treat Contact 1, 3 and 4 as equivalent and without loss of generality, set $V_1=V_3=V_4=0$. Therefore Eq. (\ref{eq6}) reduces to:
\begin{equation}
\begin{split}
I_1^\sigma&=-G_{12}^\sigma V_2\\
I_2^\sigma&=(G_{21}^\sigma+G_{23}^\sigma+G_{24}^\sigma)V_2\\
I_3^\sigma&=-G_{32}^\sigma V_2\\
I_4^\sigma&=-G_{42}^\sigma V_2
\end{split}
\label{eq7}
\end{equation}
We focus on the outputs, i.e., terminal 1, 3 and 4. According to Equation (\ref{eq7}), we only have to obtain $G_{12}^\sigma$, $G_{32}^\sigma$, and $G_{42}^\sigma$. We assume that $|V_2|$ is so small that the potential change in the scattering region induced by $V_2$ can be neglected. According to Landauer-B\"{u}ttiker formula at zero temperature, the conductance can be written as:
\begin{align}
G_{ij}^\uparrow=\frac{e^2}{h}(T_{ij}^{\uparrow \uparrow}+T_{ij}^{\uparrow \downarrow})\\
G_{ij}^\downarrow=\frac{e^2}{h}(T_{ij}^{\downarrow \downarrow}+T_{ij}^{\downarrow \uparrow})
\end{align}
where $T_{ij}^{\sigma\tau}$ is the probability of transmission between terminal $j$ and terminal $i$. When $\sigma=\tau$, $T_{ij}^{\sigma\tau}$ is the probability of transmission of electron with the spin conservation. On the other hand, when $\sigma\neq\tau$, $T_{ij}^{\sigma\tau}$ is the probability of transmission of electron with a spin flip. Considering the spin-dependent conductance, we calculate the spin polarization by the current polarization:
\begin{equation}
P_{ij}=G_{ij}^\uparrow-G_{ij}^\downarrow
\label{eq10}
\end{equation}
In the calculations, if not stated otherwise, we adopt following geometric parameters: width of each lead $W_l$=40nm, width of each strip $W_s$=40nm, radius of the inner half circle of the half ring $R$=180nm, width of the ring channel $W_r$=50nm, center of the ring $x_r$=400nm and $y_r$=0, length of the scattering region $L$=1000nm, geometric parameters of the top surface gate $l_1$=100nm, $r_1$=300nm, $b_1$=160nm, $t_1$=240nm, depth of the 2D electron gas plane $d$=50nm, lattice constant $a$=2nm, Fermi energy $E_F$=8meV. The distance between Contact 2 and 4 $d_s$ is calculated by $d_s=2(R+W_s/2)$=400nm. In our model the two surface gates are mirror-symmetrical about the $x$-axis. We choose the material parameters corresponding In$_{0.5}$Ga$_{0.5}$As, i.e., $m^*=0.0465m_0$, where $m_0$ is the electron rest mass, and $g=-8.97$. The magnetic strength is $B$=1.5T in the $-z$ direction.

\begin{figure}
  \centering
  \subfigure{
    \includegraphics[height=2in,keepaspectratio=true,angle=0]{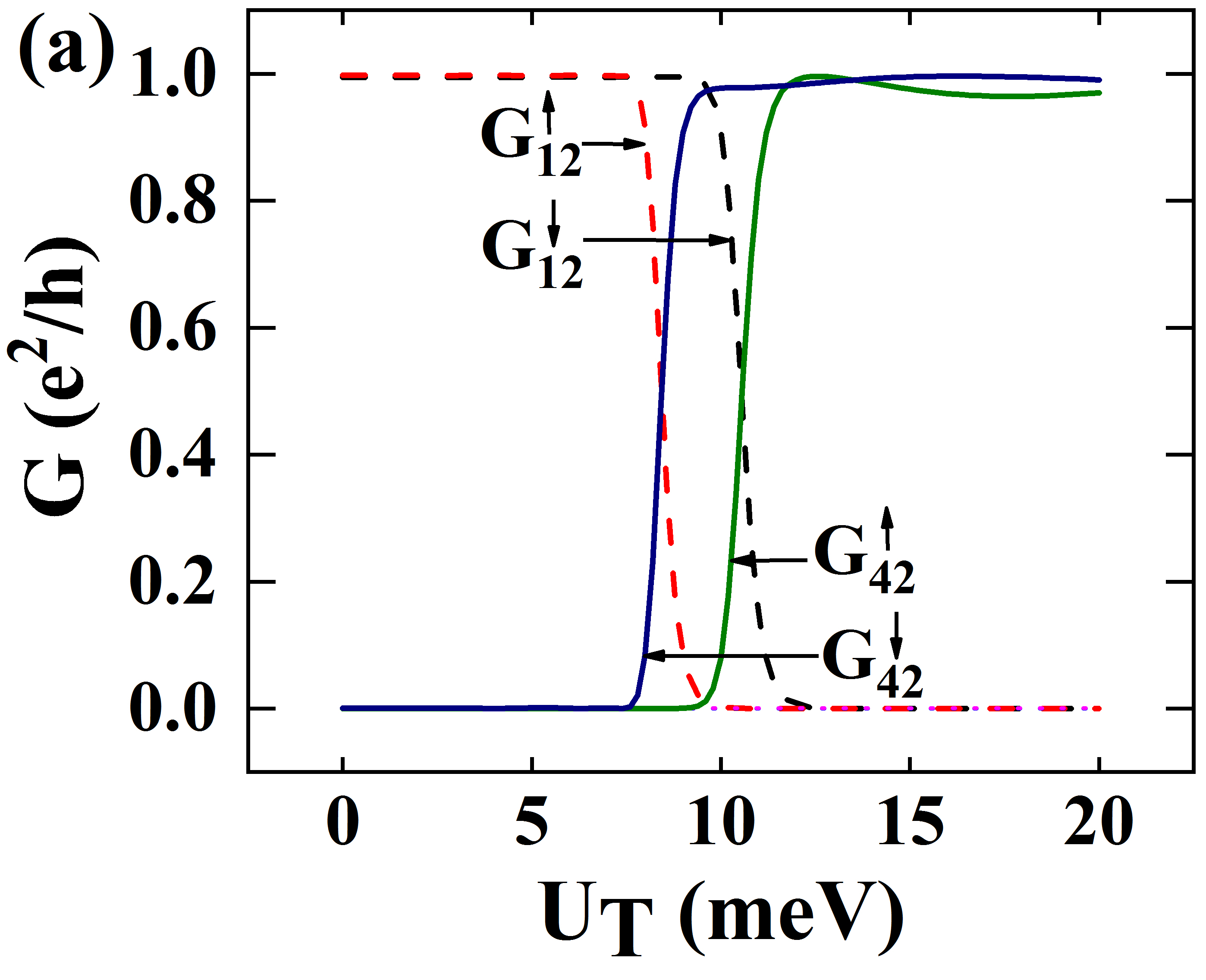}
    \label{fig2a}
  }
  \subfigure{
    \includegraphics[height=2in,keepaspectratio=true,angle=0]{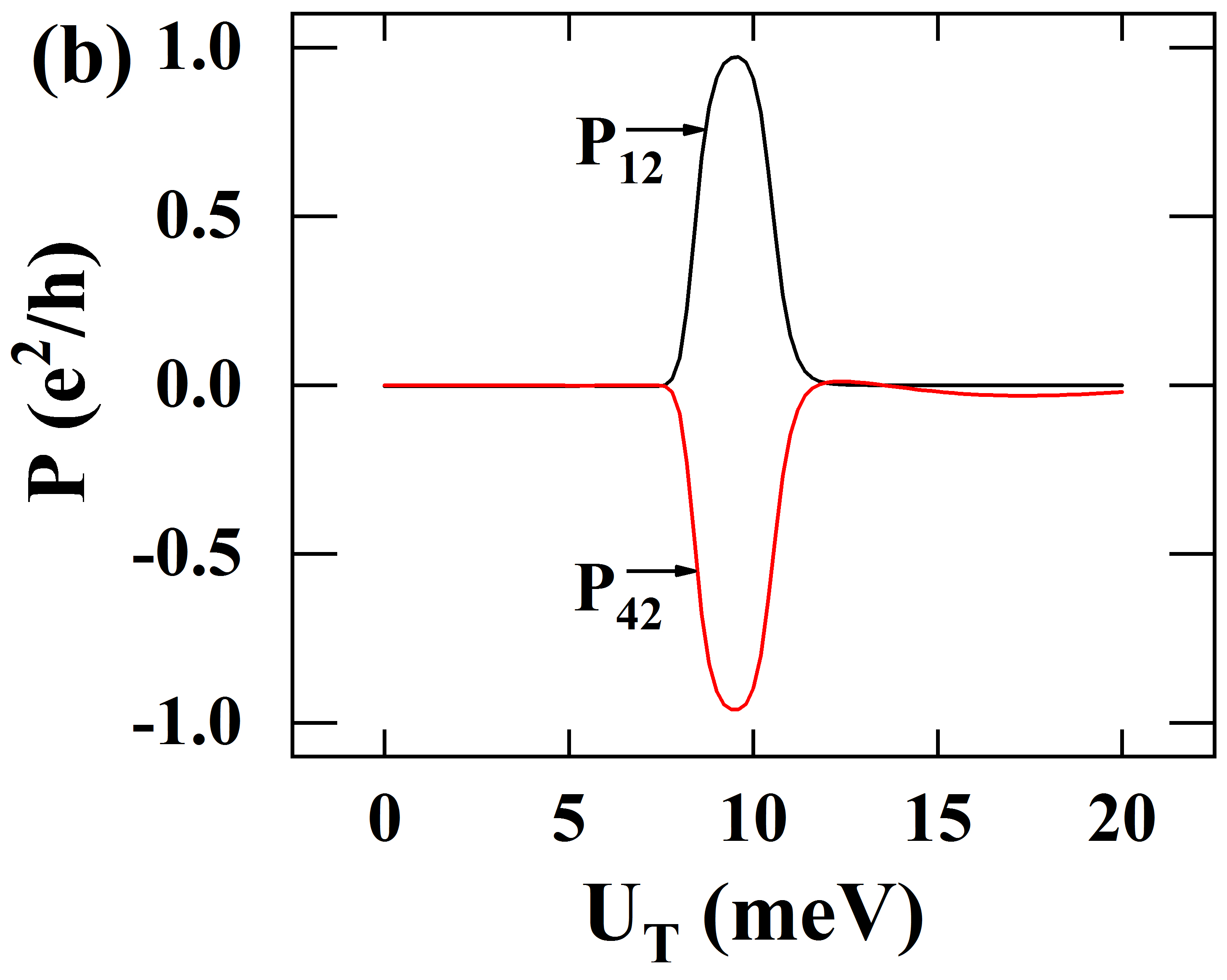}
    \label{fig2b}
  }
  \caption{(a) The conductances $G_{12}^\sigma$,  $G_{32}^\sigma$, and  $G_{42}^\sigma$ ($\sigma=\uparrow, \downarrow$) as the function of $U_T$ while the bottom gate voltage $V_{B}$ remains zero. Although not labeled, conductances $G_{32}^\uparrow$ and $G_{32}^\downarrow$ remain zero. (b) Polarizations $P_{12}$ and $P_{42}$ as the function of $U_T$ while the bottom gate voltage $V_{B}$ remains zero. The magnetic strength is $B$=1.5T in the $-z$ direction.}
\end{figure}

In this section, we present results for ballistic transport and discuss the physics of spin polarization in our nanostructure. Without loss of generality, firstly we discuss the relationship between the spin conductances and the top gate voltage $V_{T}$. For convenience, we define the potential energy $U_T=-eV_{T}$ (with a negative sign for electrons), and the potential energy $U_B=-eV_{B}$, where index $T$ denotes the top gate and $B$ denotes the bottom gate. As the spin-unpolarized current is injected from Contact 2, we present the conductances $G_{12}^\sigma$, $G_{32}^\sigma$, and $G_{42}^\sigma$ ($\sigma=\uparrow, \downarrow$) (Fig. \ref{fig2a}) and polarizations $P_{12}$ and $P_{42}$ (Fig. \ref{fig2b}) as the function of $U_T$ while the bottom gate voltage $V_{B}$ remains zero. The Fermi energy ensures that only a spin-up mode and a spin-down mode are allowed to be injected into the nanostructure.

Although not labeled in Fig. \ref{fig2a}, conductances $G_{32}^\uparrow$ and $G_{32}^\downarrow$ remain zero in the range from 0 to 20meV. We see that in a narrow $U_T$ interval, when increasing $U_T$, $G_{12}^\uparrow$($G_{12}^\downarrow$) rapidly declines from $e^2/h$ to 0. This decrease is induced by the reflecting effect of the potential barrier brought by the top gate. The threshold energy when the decrease of $G_{12}^\uparrow$ occurs, which is about 9.4meV, is larger than that when the decrease of $G_{12}^\downarrow$ occurs, which is about 7.6meV. In the same $U_T$ regime, when the conductance of spin-up electrons is still large, most of the spin-down electrons are already reflected by the potential barrier. The difference between the two threshold energies can be attributed to the splitted subband structure induce by the Zeeman term, i.e., the $g\mu_BB\sigma_z/2$ term in the Hamiltonian. The energy of spin-down subband is lower than that of spin-up subband. This leads to a difference between the probabilities of transmission of spin-up and spin-down electrons for a same potential energy $U_T$. This effect gives rise to a nonzero spin polarization as shown in Fig. \ref{fig2b}, where $P_{12}$ and $P_{42}$ are calculated by Equation (\ref{eq10}). In addition, the decrease of $G_{12}^\uparrow$($G_{12}^\downarrow$) is accompanied by the increase of $G_{42}^\uparrow$($G_{42}^\downarrow$). This simultaneity illustrates that the electrons reflected by the potential barrier brought by the top gate mostly flow to Contact 4. This effect is caused by the Lorentz force induced by the magnetic field. As the magnetic induction is in the $-z$ direction, the electronic current flow is shifted to the right boundary of the channel in the flowing direction. Thus almost all the electrons reflected flow along the inner boundary of the half ring and to the Contact 4.

We assume that the injected electronic current from Contact 2 is spin-unpolarized. While $U_B$ remains zero and $U_T$ is not zero, the currents in the Contact 1 and 4 are spin-polarized when there is a difference between the probabilities of transmission of two spins in the channel area with a potential barrier, which is due to the spin Zeeman splitting. The electrons reflected by the potential barrier mostly flow to Contact 4 due to the Lorentz force and the electrons do not come out from Contact 3. According Equation (\ref{eq7}) and the sum rule, assuming that no electrons are reflected to Contact 2, we have: $(G_{12}^\uparrow+G_{32}^\uparrow+G_{42}^\uparrow)V_2=(G_{21}^\uparrow+G_{23}^\uparrow+G_{24}^\uparrow)V_2=I_2^\uparrow=I_2^\downarrow=(G_{21}^\downarrow+G_{23}^\downarrow+G_{24}^\downarrow)V_2=(G_{12}^\downarrow+G_{32}^\downarrow+G_{42}^\downarrow)V_2$. As $G_{32}^\uparrow=G_{32}^\downarrow=0$, after simplifying the equation above, we have: $G_{12}^\uparrow+G_{42}^\uparrow=G_{12}^\downarrow+G_{42}^\downarrow\rightarrow G_{12}^\uparrow-G_{12}^\downarrow=-(G_{42}^\uparrow-G_{42}^\downarrow)$, i.e., the sign of $P_{12}$ and $P_{42}$ are opposite (see Fig. \ref{fig2b}). When $U_T$ is over 12meV and $P_{12}$ is zero, $P_{42}$ only deviates a little from zero. The reason may be that a very small part of spin-up electrons is reflected to Contact 2. This small deviation will disappear when the magnetic field is strong enough. Anyway, it is not important in our discussion. From Fig. \ref{fig2b}, we can see that in our energy range of $U_T$, $|P_{42}|$ ($|P_{12}|$) reaches a maximum when $U_T$ is approximately 9.5meV. This point corresponds to the point in Fig. \ref{fig2a} when the spin-down current is mostly reflected by the potential barrier and the spin-up current just starts to be reflected. The maximum $|P|$ ($|P_{12}|$) is close to $e^2/h$, about $0.97e^2/h$.

\begin{figure}
  \centering
  \subfigure{
    \includegraphics[height=2in,keepaspectratio=true,angle=0]{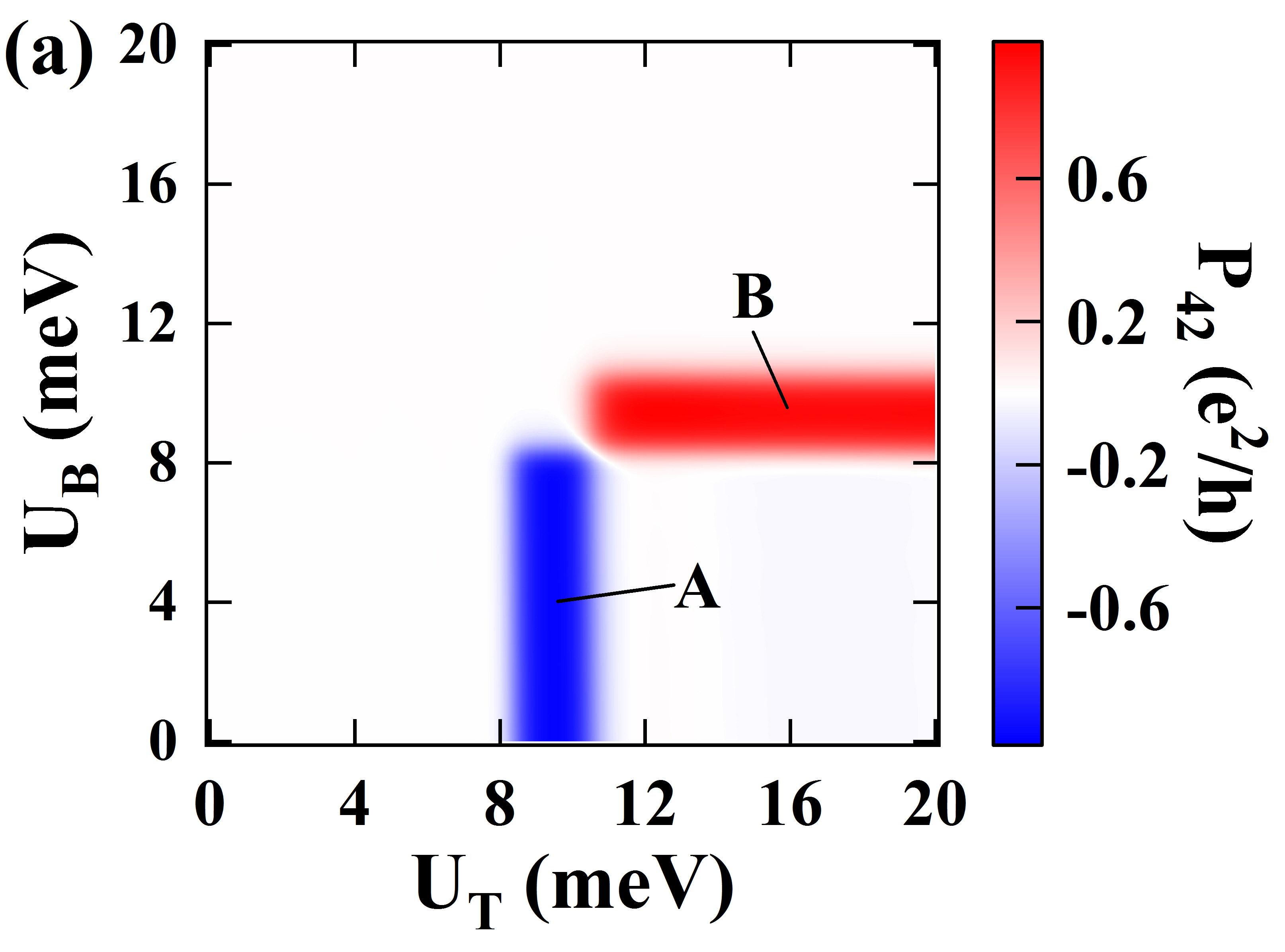}
    \label{fig3a}
  }
  \subfigure{
    \includegraphics[height=2in,keepaspectratio=true,angle=0]{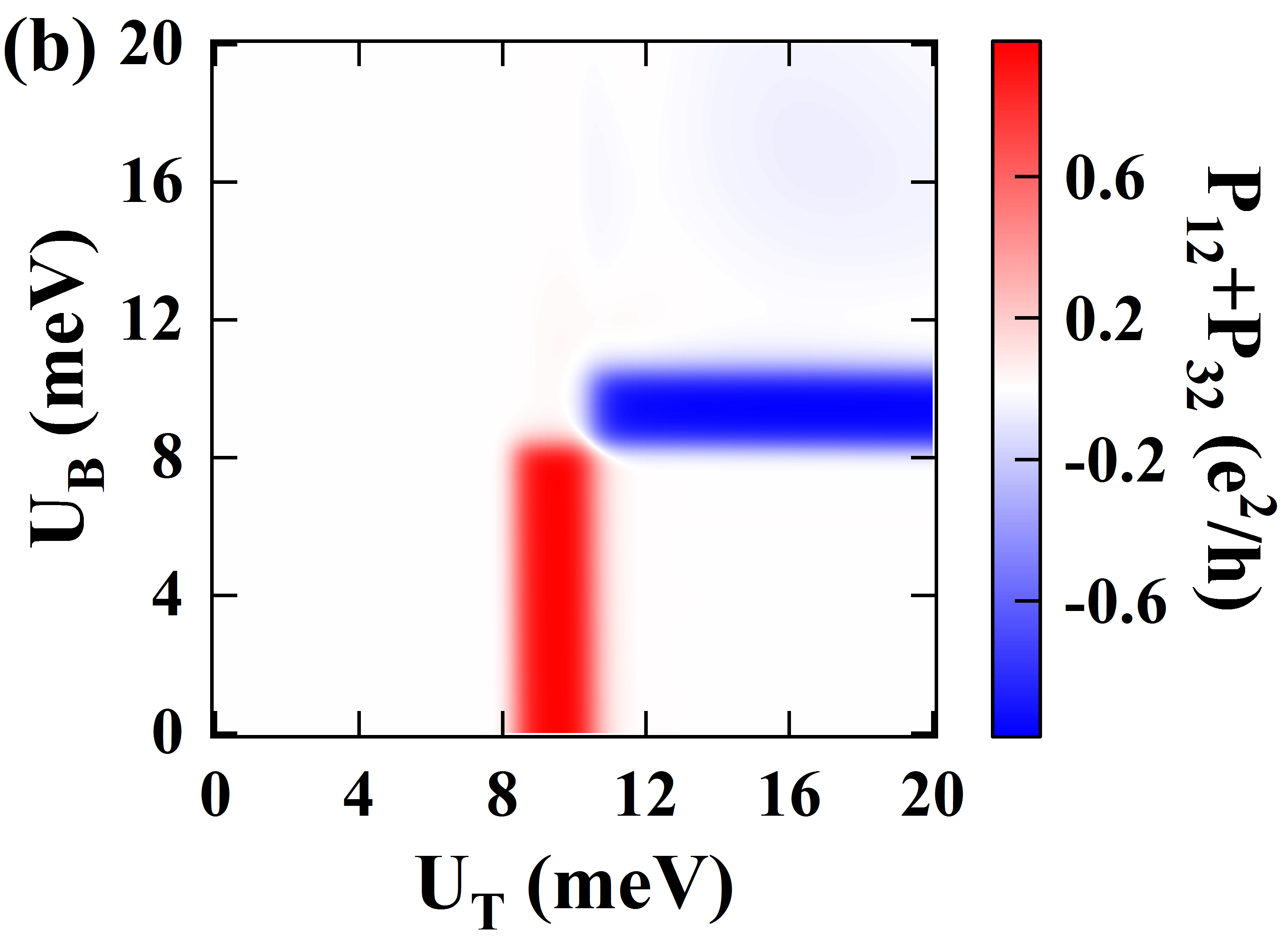}
    \label{fig3b}
  }
  \subfigure{
    \includegraphics[height=2in,keepaspectratio=true,angle=0]{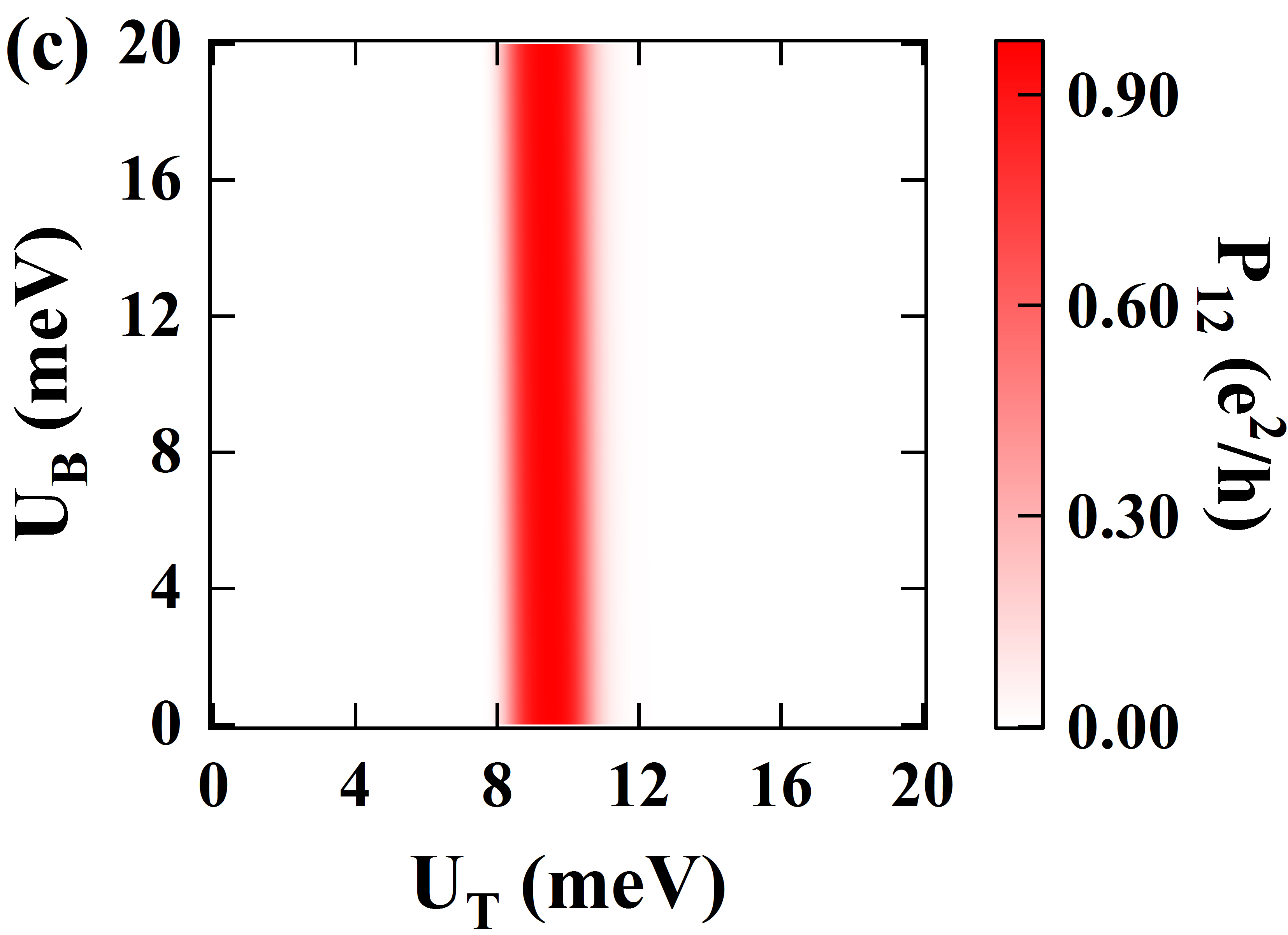}
    \label{fig3c}
  }
  \subfigure{
    \includegraphics[height=2in,keepaspectratio=true,angle=0]{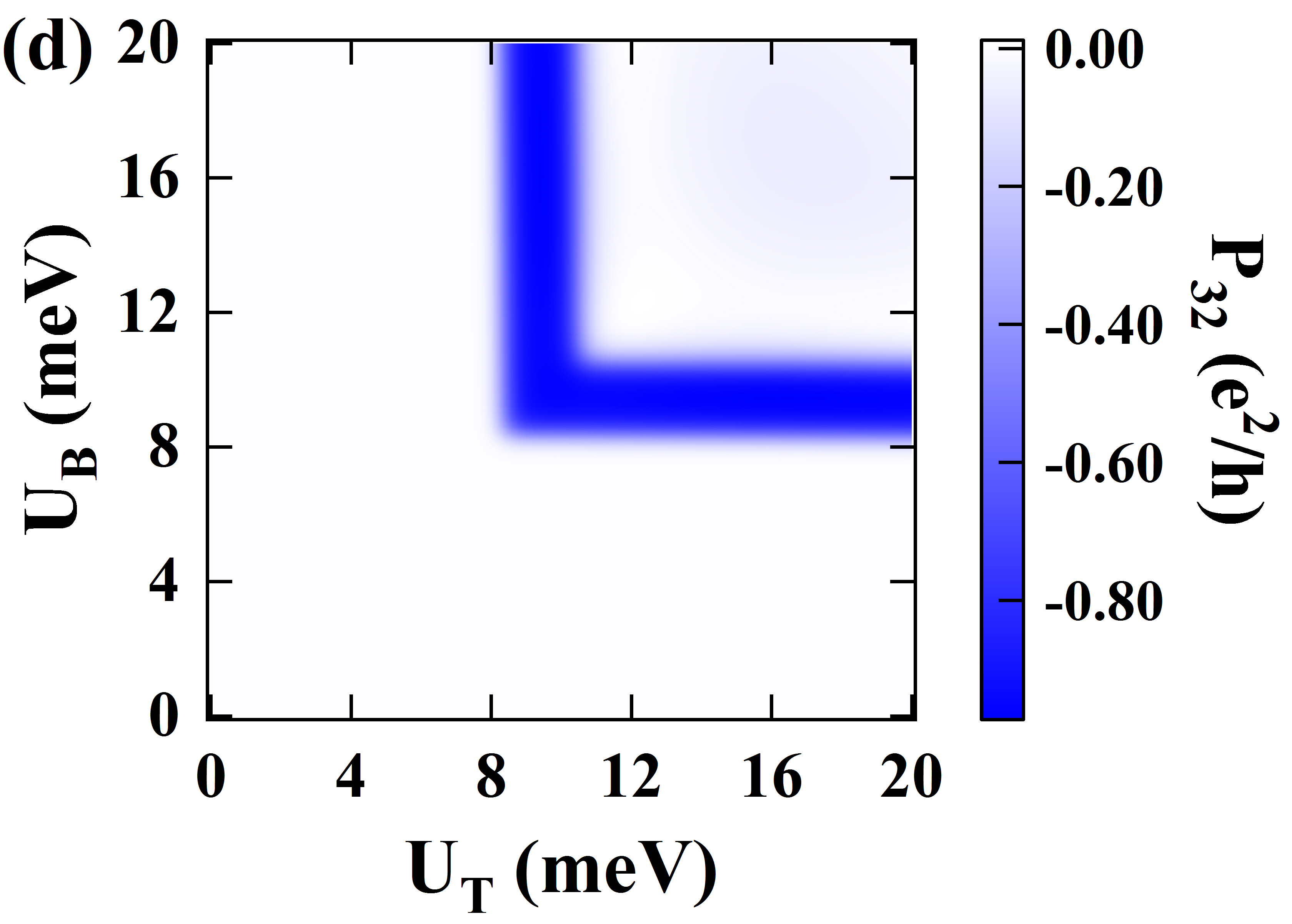}
    \label{fig3d}
  }
  \caption{(a) The conductance polarization $P_{42}$ as the function of potential energy $U_T$ and $U_B$. (b) The sum of conductance polarization $P_{12}$ and $P_{32}$ as the function of potential energy $U_T$ and $U_B$. (c) The conductance polarization $P_{12}$ as the function of potential energy $U_T$ and $U_B$. (d) The conductance polarization $P_{32}$ as the function of potential energy $U_T$ and $U_B$. The magnetic strength is $B$=1.5T in the $-z$ direction.}
\end{figure}
While $U_B$ remains zero, the sign of $P_{42}$ remains unchanged in our energy range of $U_T$. With the exploit of both the two gates, the sign of $P_{42}$ can be reversed. Fig. \ref{fig3a} shows $P_{42}$ as the function of both $U_T$ and $U_B$. We can see that $P_{42}$ is divided into three regions: a negative region, a positive region and the rest region where $P_{42}$ is about zero. Around the boundary between regions, $P_{42}$ is changed sharply with varying $U_T$ or $U_B$. As the bottom gate is geometrically equivalent to the top gate, the effect of the bottom gate on the spin current in the bottom channel is similar. The electrons reflected by the potential barrier brought by the bottom gate flow to the Contact 3 along the lower boundary of the bottom horizontal channel due to the Lorentz force. The shape of the negative region is formed by the collective effect of the top and bottom gates. The essence is that the potential energy $U_T$ is in a range where the reflected spin-down electrons exceeds the reflected spin-up electrons, while the potential energy $U_B$ must not exceed 9.5meV so as not to entirely reflect spin-down electrons by the potential brought by the bottom gate. Similarly, the positive region is formed when the potential energy $U_T$ exceeds the threshold energy when the spin-up electrons start to be reflected so as to flow to Contact 4 and the potential energy $U_B$ is in the same range where the reflected spin-down electrons exceeds the reflected spin-up electrons.

We notice that the electrons flow through the nanostructure without spin-flip incidents, but the total spin of electrons coming out from Contact 4 can be reversed by changing $U_T$ and $U_B$. These electrons originate from the injector - Contact 2. Electrons cannot be reflected to Contact 2 due to the Lorentz force. As electrons injected from Contact 2 are spin-unpolarized, according to sum rule, the total spin of electrons coming out from Contact 1 and 3 must be complementary to the spin of electrons coming out from Contact 2. This will lead to $P_{42} = -(P_{12}+P_{32})$. We present the sum of $P_{12}$ and $P_{32}$ in Fig. \ref{fig3b}. We can see that the calculation results are consistent with our derivation. We also present the results for $P_{12}$ and $P_{32}$ alone in Fig. \ref{fig3c} and Fig. \ref{fig3d}. We can see that in Contact 1 and 4 alone, the total spin cannot be reversed.
\begin{figure}[h]
\centering
\subfigure{
	\includegraphics[height=2in,keepaspectratio=true,angle=0]{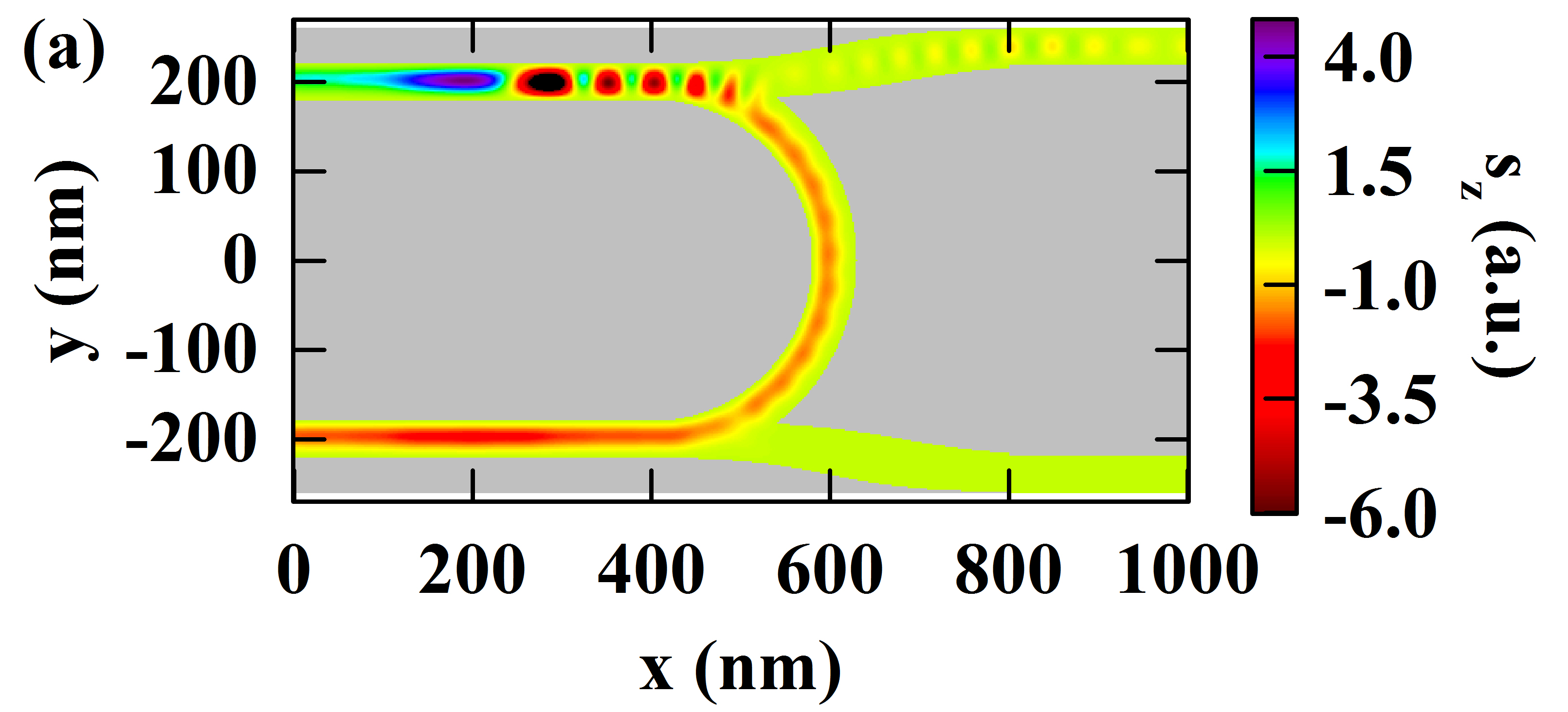}
	\label{fig4a}}
\subfigure{
	\includegraphics[height=2in,keepaspectratio=true,angle=0]{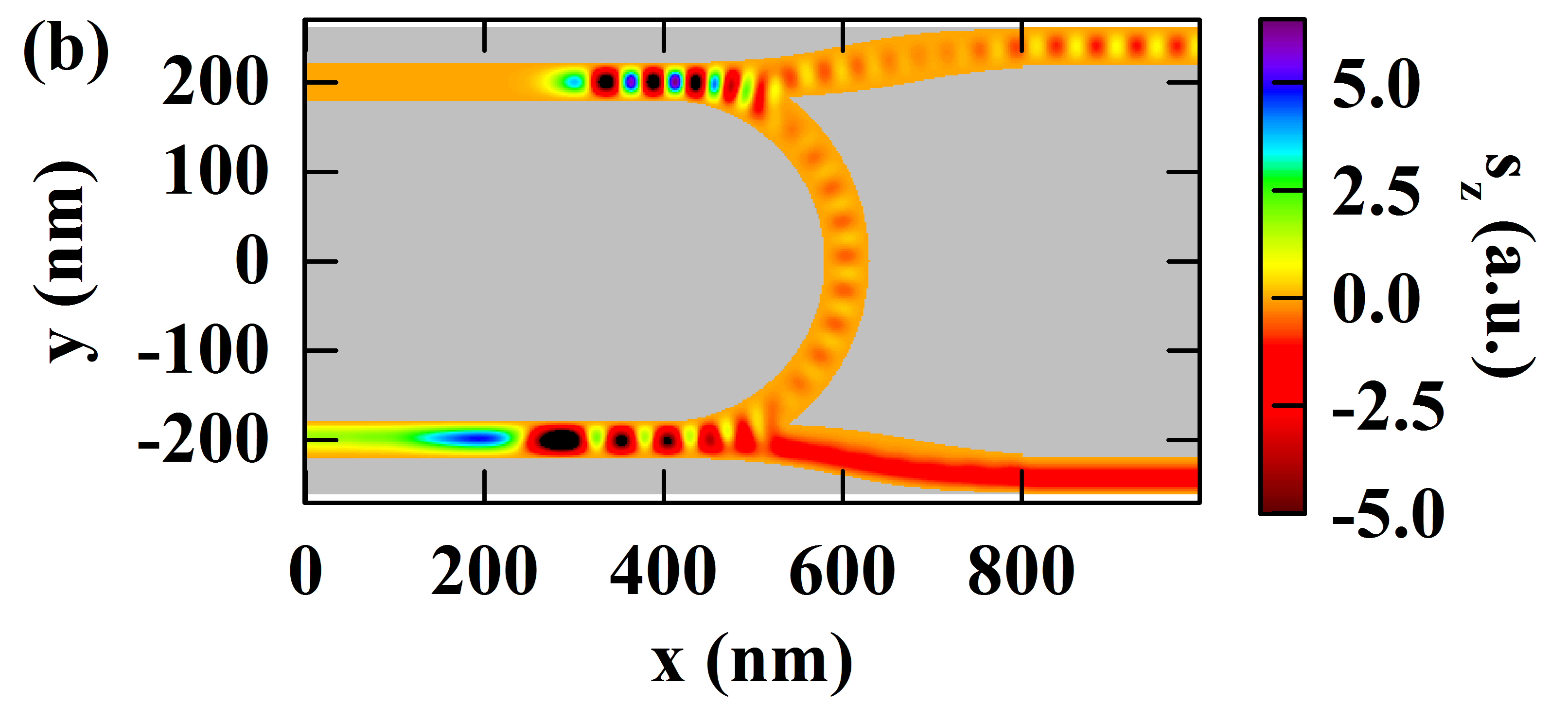}
	\label{fig4b}}
\caption{(a) The total spin density $s_z$ in the nanostructure for point A in Fig. \ref{fig3a}. (b) The total spin density $s_z$ in the nanostructure for point B in Fig. \ref{fig3a}. The magnetic strength is $B$=1.5T in the $-z$ direction.}
\end{figure}

To illustrate the electronic current, we calculate the total spin density $s_z$ (Fig. 4) in the nanostructure for point A and B in Fig. \ref{fig3a}. The $s_z$ of spin-up electrons is positive while the $s_z$ of spin-down electrons is negative. We can see that electrons injected from Contact 2 are closer to the right boundary of the channel due to the external magnetic field. At point A, $P_{42}\approx-0.96$. Most of the spin-up electrons come out of the structure from Contact 1, while most of the spin-down electrons are reflected by the potential barrier underneath the top gate (Fig. \ref{fig4a}). The spin-down electrons flow to the bottom channel along the right boundary of the structure in the flowing direction and come out from Contact 4. At point B, $P_{42}\approx0.94$. Electrons cannot flow through the potential barrier underneath the top gate and are entirely reflected to the bottom channel. Again the spin-down electrons are mostly reflected by the potential barrier brought by the bottom gate and flow to Contact 3, while the spin-up electrons can flow through the potential and come out from Contact 4 (Fig. \ref{fig4b}).


In summary, we design a four-terminal nanostructure with two surface metal gates and show that it can serve as a spin splitter/switcher. In the ballistic transport regime, we utilize the gate-controlled potential barriers, combined with Zeeman splitting effect and edge current to control the spin freedom of the outputs. We show that by tuning the voltage of gates, the injected spin-unpolarized current can be split into different spin current with a high efficiency, and flow out of the geometry from different terminals. Furthermore, by tuning the voltage of gates, the outflow spin of one terminal can be switched.

 This research was supported by the National Key R\&D Program of China under Grant No. 2016YFF0200403
  and the Key Program of National Natural Science Foundation of China under Grant No. 11234009.


\pagebreak

%
%
%
%
%
%
%
%

\end{document}